%% file: main.tex
\documentclass[a4paper]{article}
\usepackage{hyperref}
\usepackage{tikz}
\usepackage{pgfplots}
\usepackage[skip=5pt]{caption}
\usepackage{lineno}
\newcommand{\etal}{\textit{et al}.}
\newcommand{\figcap}{\setlength{\baselineskip}{.5em}}

\usepackage{INTERSPEECH2018}

\title{Modeling Singing F0 With Neural Network Driven Transition-Sustain Models}
\name{Kanru Hua}
\address{University of Illinois, U.S.A.}
\email{khua5@illinois.edu}

\begin{document}

\maketitle

\begin{abstract}
This study focuses on generating fundamental frequency (F0) curves of singing voice from musical scores stored in a midi-like notation. Current statistical parametric approaches to singing F0 modeling meet difficulties in reproducing vibratos and the temporal details at note boundaries due to the oversmoothing tendency of statistical models. This paper presents a neural network based solution that models a pair of neighboring notes at a time (the transition model) and uses a separate network for generating vibratos (the sustain model). Predictions from the two models are combined by summation after proper enveloping to enforce continuity. In the training phase, mild misalignment between the scores and the target F0 is addressed by back-propagating the gradients to the networks' inputs. Subjective listening tests on the NITech singing database show that transition-sustain models are able to generate F0 trajectories close to the original performance.
\end{abstract}

\section{Introduction}

The problem of automatically modeling singing expressions has received growing interests in recent years. Of the various parameters being considered, the fundamental frequency (F0) trajectory carries both melodic information and details related to the perceived naturalness and singing styles.

A comprehensive overview on singing expression control by Umbert \etal\cite{umbert-2015} has classified fully automatic F0 modeling methods into three categories: rule-based, statistical modeling, and unit selection. In particular, HMM-based statistical parametric systems \cite{oura-2010, oura-2012, saino-2010} have been extensively developed to model both singing timbre and F0 features. The common setup for HMM-based singing models was derived from speech synthesis systems \cite{zen-2009} wherein each phoneme is divided into a certain number of states associated with probability distributions of the feature vectors; alternatively the states can be assigned by subdividing different portions of a musical note \cite{saino-2010}. The generation of continuous F0 trajectories from a discrete state-space requires a maximum-likelihood estimation under continuity constraints and in such a process, temporal details are often lost \cite{toda-2007}. Moreover, the discrete assignment of states sets a smoothness assumption that prohibits efficient characterization of periodic structures such as vibratos.

The deep neural network (DNN)-based singing synthesis system in \cite{nishimura-2016} can be thought of as a drop-in enhancement to the HMM-based system in that the statistics of feature vectors are predicted from musical context vectors using a feed-forward DNN. Although the enhanced system predicts feature vectors on a frame-by-frame basis (as opposed to discretizing the timing into states), the maximum-likelihood parameter generation step is still required to reduce parameter discontinuities at phoneme and note boundaries. While the use of more sophisticated architectures such as recurrent \cite{ozer-2015} and autoregressive convolutional \cite{blaauw-2017} neural networks eliminates the need for a parameter generation step, it becomes unclear on how to intuitively control different performance-related parameters (e.g. note timing and vibrato frequency).

There also has been an active development in rule-based \cite{ardaillon-2015} and unit selection \cite{umbert-2013} singing expression modeling systems, both were shown to generate realistic and detail-rich F0 trajectories (after some parameter tuning). The two methods are similar in that the F0 is decomposed into the weighted sum of overlapping segments, wherein each segment is either defined by a B-spline basis or an expression database. The present study inherits the said segment-wise modeling strategy, but generates note-pair segments using a feed-forward neural network similar to \cite{nishimura-2016}. Inspired by \cite{ozer-2015}, we use a second network (which is also feed-forward instead of the LSTM-RNN in \cite{ozer-2015}) to generate vibrato segments.

In the rest of this paper, an overview of the proposed method is first given in section 2. Then the note transition model and the sustain model are respectively described in section 3 and section 4. The training strategy and results from a subjective listening test are summarized in section 5. Finally, section 6 concludes the study with additional notes on future studies.

\section{System Overview}

\begin{figure}[htb]
  \centerline{\includegraphics[width=6cm]{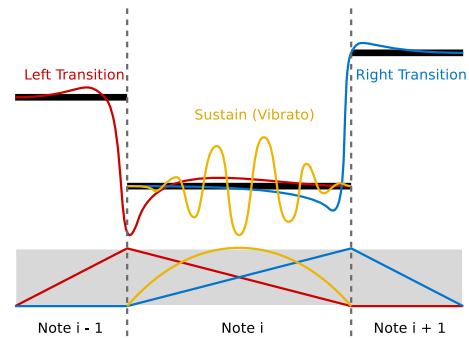}}
  \caption{\figcap{A diagram showing the decomposition of a F0 trajectory into dinotes (left and right) and mononotes (middle), weighted by the envelopes in the gray region.}}
  \label{fig:schematics}
  \vspace{-10pt}
\end{figure}

In the proposed singing F0 modeling system, musical notes are the elementary units of the formulation. With references to figure~\ref{fig:schematics}, to facilitate the following description, we define a pair of notes covering the transition region between note $i-1$ and note $i$ as a \textit{dinote} $D_{i-1, i}$, similar to the notion of a diphone in concatenative speech synthesis. In contrast, the sustained region within a note that may contain vibrato is defined as a \textit{mononote} $M_i$. The F0 trajectory of a musical score containing $N$ notes can thus be decomposed into the weighted sum of $N-1$ dinotes and $N$ mononotes.

The log F0 trajectory within a dinote is generated by a feed-forward neural network based \textit{transition model} and correspondingly, a \textit{sustain model} generates the time-varying vibrato depth parameter in a mononote; the vibrato depth is converted to log F0 deviation by amplitude-modulating a sinusoid signal. To ensure the continuity of the final trajectory, the dinotes are modulated by linear cross-fading envelopes \cite{umbert-2013} and the mononotes are modulated by a duration-determined convex parabola prior to summation (see the bottom of Figure~\ref{fig:schematics}),
\begin{align}
  \log f_0(t) &= \sum_{i = 1}^N D_{i-1, i}(t) {w_D}_{i-1, i}(t) + \sum_{i = 1}^N M_i(t) {w_M}_i(t)
  \label{eq:weighted-sum}
\end{align}
where
\begin{align}
  {w_D}_{i-1, i}(t) &= \begin{cases}
    0 & t < T_{i-1}\ \mathrm{or}\ t > T_{i+1} \\
    \frac{t - T_{i-1}}{T_i - T_{i-1}} & T_{i-1} \leq t < T_i \\
    \frac{T_{i + 1} - t}{T_{i + 1} - T_i} & T_i \leq t \leq T_{i + 1}
  \end{cases} \\
  {w_M}_i(t) &= \begin{cases}
    0 & t < T_{i}\ \mathrm{or}\ t > T_{i+1} \\
    \frac{4 (t - T_i)(T_{i+1} - t)}{(T_{i+1} - T_{i})^2} & T_i \leq t \leq T_{i + 1}
  \end{cases} \\
  D_{0,1}(t) &= D_{1,2}(0)
\end{align}
and $T_{i}$ is the onset of note $i$. The generation of $D_{i-1, i}$ and $M_i$ depends on the musical context. Note that there is no need to extract dinote and mononote segments prior to training as the separation can be learned implicitly when using a multi-stage training method directly optimizing the final log F0 trajectory $\log f_0(t)$. The said process is discussed in detail in the following sections.

\section{Modeling Note Transitions}
\label{sec:tran-model}

The transition model is a direct extension of the phoneme-level DNN in \cite{nishimura-2016} to dinote units. With details listed in Table~\ref{tab:tran-model-inputs}, given the musical and phonetic context of a pair of neighboring notes and a relative time position within the dinote, the model predicts the F0 at the specified position. The complete trajectory covering a dinote can thus be generated by iterating the model over all frames in the dinote.
\begin{table}[h]
\centering
\caption{\figcap{Inputs to the transition model.}}
\begin{tabular}{|l|c|c|}
  \hline
    Name & Type & Size \\ \hline
    Duration of the first note & Real & 1 \\ \hline
    Duration of the second note & Real & 1 \\ \hline
    Pitch interval of the transition & Real & 1 \\ \hline
    The first note being silent & Binary & 1 \\ \hline
    The second note being silent & Binary & 1 \\ \hline
    Type of the second note's onset & One-hot & 3 \\
    (legato / vowel-leading / default) & & \\ \hline
    Position relative to the note boundary & Real & 1 \\ \hline
\end{tabular}
\label{tab:tran-model-inputs}
\vspace{-5pt}
\end{table}

To improve the model's generalization across pitch ranges, for each dinote section the pitch of the second note is subtracted from the target log F0 during training. This can be alternatively implemented as a residual connection (from the input) adding a note-dependent offset to the network's output, thereby allowing direct gradient evaluation on the weighted sum in equation (\ref{eq:weighted-sum}).

\begin{figure}
\centering
\input{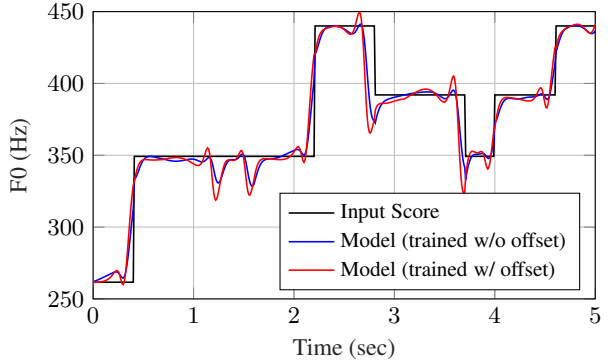}
\caption{\figcap{Comparing the F0 synthesized by a transition model trained with back-propagation-based offset correction to the baseline transition model.}}
\label{fig:offset-correction}
\vspace{-10pt}
\end{figure}

Another issue observed in the training process is the slight timing mismatches between musical scores and the singing voice in the database due to the imprecise nature of human singing. In statistical parametric systems, naive attempts to train the model on non-systematically time-offsetted data using a unimodal cost function (e.g. Euclidean distance) inevitably leads to an average F0 curve with reduced level of details. Although the use of a mixture density network \cite{zen-2014} was shown to alleviate the oversmoothing problem, our preliminary tests indicated that simply realigning the musical score with the recordings is also a highly effective solution. In order to automate the realignment process, the network's gradient against the input relative time position is taken and the note onsets are trained along with the transition model. The maximum offset correction is limited within $\pm 30$ milliseconds to prevent the build-up of occasional realignment errors. Figure~\ref{fig:offset-correction} shows the results from transition models trained with and without back-propagating gradients to the input offsets on a 5-second excerpt from the test data. It is seen that offset-corrected training produces more defined pitch ornamentations near note boundaries.

\section{Modeling Vibratos}

The experiments in \cite{ardaillon-2015, maher-1990} suggested that a constant-frequency sinusoid with proper amplitude fading on the ends gives a perceptually accurate approximation to singing vibrato; hence the neural network in the sustain model only needs to predict the vibrato depth from the mononote musical context. However, without explicitly addressing the phase and frequency variation across time and across different occurrences of vibratos, it becomes hard to fit equation (\ref{eq:weighted-sum}) on training data and, an underfit model will produce damped vibrato depths. The problem is further complicated by the fact that a high degree of randomness was observed in vibrato phase (or the onset position and polarity of frequency oscillations).

Under the assumption that vibrato phase and small variations in the frequency are perceptually irrelevant, we use a method similar to addressing the timing issues in section~\ref{sec:tran-model} that supplements the network with note-dependent phase-adjusting input parameters. These auxillary inputs receive back-propagated gradients during training and are set to their default values when evaluating the models on unseen data. The mononote trajectory $M_i(t)$ can then be expressed as a sinusoid modulated by vibrato depth prediction $A_i(t)$,
\begin{align}
  M_i(t) &= A_i(t) \sin(\omega_i \varphi_i[(t - T_i) / d_i] d_i + \theta_i)
\end{align}
where $d_i = T_{i+1} - T_i$, that is, the duration of the i-th note; $\omega_i$ is a pre-determined angular frequency; $\theta_i$ is a trainable parameter for the phase shift; $\varphi_i[\cdot]$ is a trainable warping function on $[0, 1]$ that models the time-varying frequency disturbance. Essentially, the combination of $\omega_i$, $\varphi_i$ and $\theta_i$ defines a wide range of time translation and non-linear stretching operations on the sine function, whereas $A_i(t)$ provides an amplitude envelope. Inspired by the phase interpolation technique for sinusoidal models \cite{quatieri-1986}, we implement $\varphi_i$ using a cubic Hermite spline function with fixed end points at $(0, 0)$ and $(1, 1)$,
\begin{align}
  \varphi_i(\tau) &= -2\tau^3 + 3\tau^2 + \nonumber \\
    &\phantom{=}\quad (\tau^3 - 2\tau^2 + \tau) \alpha_i + (\tau^3 - \tau^2) \beta_i,\ 0 \leq \tau \leq 1
\end{align}
where $\alpha_i$ and $\beta_i$ are the slope parameters with default value $1$. For $\varphi_i$ to be a valid warping function, its derivative on the range of definition should be positive and a sufficient condition is that both $\alpha_i$ and $\beta_i$ are between $0$ and $3$; in practice their values are restricted within $[0.5, 2.0]$ to prevent extreme stretching.

The training of mononote trajectory $M_i(t)$ relies on a pre-determined vibrato frequency $\omega_i$ and a good initial value for vibrato phase parameter $\theta_i$. We estimate both $\omega_i$ and $\theta_i$ by first training the transition model and then perform peak-picking \cite{quatieri-1986} on the FFT spectra of the residual F0 trajectory, that is, the difference between the target log F0 and the one re-synthesized from the transition model only.

Finally, the input parameters to the vibrato depth prediction network $A_i(t)$ are listed in Table~\ref{tab:sus-model-inputs}.

\begin{table}[h]
\centering
\caption{\figcap{Inputs to the sustain model (for $A_i(t)$ only).}}
\begin{tabular}{|l|c|c|}
  \hline
    Name & Type & Size \\ \hline
    Duration of the mononote & Real & 1 \\ \hline
    The previous note being silent & Binary & 1 \\ \hline
    The next note being silent & Binary & 1 \\ \hline
    Pitch of the mononote & Real & 1 \\ \hline
    Pitch difference from the previous note & Real & 1 \\ \hline
    Pitch difference from the next note & Real & 1 \\ \hline
    Type of the mononote's onset & One-hot & 3 \\
    (legato / vowel-leading / default) & & \\ \hline
    Position in the mononote (forward) & Real & 1 \\ \hline
    Position in the mononote (backward) & Real & 1 \\ \hline
\end{tabular}
\label{tab:sus-model-inputs}
\vspace{-5pt}
\end{table}

\subsection{The Training Procedure}
\label{sec:training}

At this point the topology and composition of the singing F0 models have been described in great detail. The remaining concept yet has to be clarified is how to train the transition and sustain models from a dataset of parallel musical scores and recorded vocals. We use a multi-stage gradient descent strategy that first trains the transition model and then switches to the sustain model to avoid unwanted interactions that may cause overfitting (e.g. vibratos being used for modeling note transitions). It was also experimentally found that L1 loss on log F0 better fits the vibrato than L2 loss. With the use of Adam optimizer at an $10^{-3}$ learning rate (unless otherwise noted), the detailed procedure is listed as the following.
\begin{enumerate}
  \item Disable the sustain model. Train the transition model without offset correction for 10 epochs;
  \item Turn on offset correction for the transition model and train for 10 epochs;
  \item Re-synthesize the F0 trajectories using the trained transition model and compute the residual F0; estimate vibrato frequency and initial phase parameters from the residual;
  \item Disable the transition model and enable the sustain model. Train the sustain model with phase correction for 60 epochs;
  \item Re-enable the transition model and train both models for 10 epochs at an exponentially decreasing rate with decay constant $0.75$.
\end{enumerate}
Since offset and phase corrections are not performed on the test data, the average test loss cannot be used as an early stop criterion. The procedure listed in this paper thus fixes the number of epochs for each stage. An objective measure for test-set performance is yet to be investigated.

\section{Experiments}

The proposed transition-sustain models were trained on 21 songs from the \texttt{nitech\_jp\_song070\_f001} singing database\footnote{The full database consists of 70 songs, 31 of which can be accessed from \url{http://hts.sp.nitech.ac.jp/}.} and were evaluated on the rest 10 songs. The database consists of Japanese children's songs by a female singer and has been previously used in the evaluation of HMM \cite{oura-2010} and DNN-based \cite{nishimura-2016} singing synthesis systems. The neural networks for both transition and sustain models have 2 hidden layers each with 64 units and are trained according to section~\ref{sec:training}.

\begin{figure}
  \centerline{\includegraphics[width=6cm]{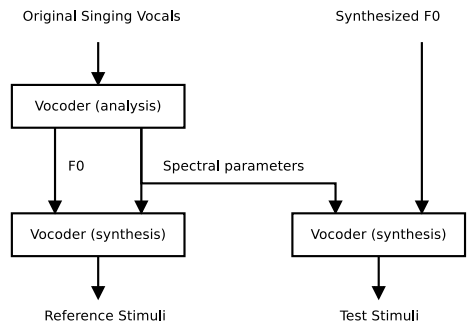}}
  \caption{\figcap{The method for generating pairs of test stimuli with original and synthesized F0 by modifying the original singing.}}
  \label{fig:test-procedure}
  \vspace{-5pt}
\end{figure}

A subjective test following the ITU-R BS. 1116 \cite{itur-2015} quality impairment evaluation method was carried out to compare the proposed method with the HMM-based Sinsy system \cite{oura-2010} and the B-spline-based multi-layer F0 model \cite{ardaillon-2015}. As recommended in the paper, parameters for the multi-layer F0 model were tuned by hand on the training data to fit the overshoots, undershoots and vibratos in the original performance. Towards the goal of measuring how well these models replicate the singer's style on test data, we modified the original vocals by replacing the F0 with the synthesized version (see Figure~\ref{fig:test-procedure} for the flowchart) and evaluated the difference between modified singing and the original version in the listening test. To eliminate potential biases due to audio defects introduced by the vocoder, the original recordings were also processed by the vocoder but without parameter modification. For the said purpose, a high-quality full-band harmonic-noise model vocoder\footnote{\url{https://github.com/Sleepwalking/libllsm2}} was used in this experiment.

The listening test was conducted using an online interface \cite{cartwright-2016} on 10 short excerpts randomly chosen from the test set, each lasting between 5 to 10 seconds. Prior to the test, subjects were advised to use a headphone. The music excerpts were rendered using all systems being tested, yielding a total of 30 stimuli in groups of 3. Each time, the listener is presented with a known reference (the original singing processed by the vocoder) and a randomly chosen stimulus paired with a hidden reference identical to the known reference. After listening to the known reference, the subjects were asked to identify the stimulus with modified F0 and give a comparative rating on a continuous scale between 1 and 5 based on the following criterion \cite{itur-2015}. Looping and switching between stimuli and the reference were allowed.
\begin{table}[h]
\centering
\caption{\figcap{Five-grade impairment scale.}}
\begin{tabular}{|l|c|}
  \hline
    Impairment & Grade \\ \hline
    Imperceptible & 5 \\
    Perceptible, but not annoying & 4 \\
    Slightly annoying & 3 \\
    Annoying & 2 \\
    Very annoying & 1 \\ \hline
\end{tabular}
\label{tab:bs1284}
\vspace{-5pt}
\end{table}

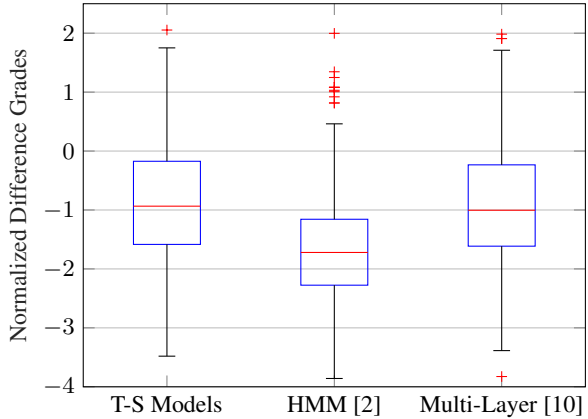
\begin{figure}
\centering
\input{boxplot.tikz}
\caption{\figcap{A box plot of the listening test results comparing three F0 modeling systems with reference to the original singing. The boxes cover the IQR ($25\%$ to $75\%$ quantile)} and the red lines are placed at the medians. Outliers (more than 3 IQRs from the median) are marked in red ``+".}
\label{fig:boxplot}
\vspace{-10pt}
\end{figure}

According to the post-test survey, of the 32 listening test participants, 7 are experts in speech and music-related research; 21 have experience (2 years or more) in either vocal performance or music production; 16 are native or fluent Japanese speakers. Breakdown analysis of test results shows no significant difference due to different levels of experience in research, music or language education. During data analysis, data points from 2 subjects were removed as their responses did not meet the objective criteria for listeners' expertise as specified in \cite{itur-2015}. The differences between hidden reference ratings and F0-modified stimuli ratings, after normalizing across listeners, are computed from the remaining data and visualized in Figure~\ref{fig:boxplot}.

From the box plot it can be seen that the proposed method performs significantly better than the HMM-based system in replicating the singing style, but the distribution of ratings are almost identical between the proposed method and the multi-layer F0 model, with both systems having a median difference grade around $-1$ that corresponds to the ``perceptible, but not annoying" impairment grade. The test results are supplemented by Table~\ref{tab:stats} which lists the median time for a rating to be made on each pair of stimuli and the accuracy of identifying the hidden reference. Listeners tend to spend 6 more seconds (which happens to be the mean duration of a stimulus) on telling apart transition-sustain models/multi-layer model from the original F0 compared to the time spent on the HMM-based model. In $15\%$ cases the F0 generated by transition-sustain models/multi-layer model successfully spoofed the listener to mistake it as the hidden reference. The results lead us to believe that both transition-sustain models and the multi-layer F0 model are able to generate F0 expressions resembling the original performance to an extent that makes comparison between the two methods difficult. The advantage of the proposed method lies in being fully data-driven, while the multi-layer F0 model requires hand-tuning and its automation is still under investigation \cite{ardaillon-2016}.

\begin{figure}
\centering
\input{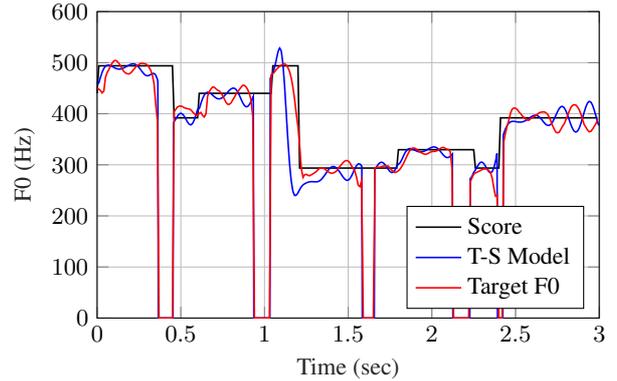}
\caption{\figcap{F0 generated by the transition-sustain model versus the target performance on the test stimulus that received the lowest average rating. A timing error is observed at $1.2$ sec.}}
\label{fig:bad-example}
\vspace{-10pt}
\end{figure}

Figure~\ref{fig:bad-example} shows an example of a F0 trajectory generated by the purposed method overlaid on the input score and the original F0. On this specific example transition-sustain models received the lowest average rating. By listening to the F0-modified stimulus, we attribute the problem to a timing error at $1.2$ sec where the note transition starts $50$ milliseconds earlier than the original performance. The early transition is possibly caused by realignment errors during transition model training (section~\ref{sec:tran-model}).

\begin{table}
\centering
\caption{\figcap{Supplementary listening test statistics.}}
\begin{tabular}{|c|c|c|c|}
  \hline
    & T-S Models & HMM & Multi-Layer \\ \hline
    Median decision & $33.2$ & $26.3$ & $33.4$ \\
    time (s) & & & \\ \hline
    Identification rate & $84.3\%$ & $93.3\%$ & $85.0\%$ \\ \hline
\end{tabular}
\label{tab:stats}
\vspace{-5pt}
\end{table}

\section{Conclusions}

This paper proposed a singing F0 modeling method using separate feed-forward neural networks for note transitions and vibratos. Without relying on a parameter generation or smoothing algorithm, F0 trajectories can be generated directly from the weighted sum of neural network outputs. The separate modeling of baseline melodic component and vibrato component allows intuitive control of pitch ornamentations by the user.

It has also been shown that oversmoothing problems can be addressed by back-propagating the gradients to the inputs for both transition and vibrato depth prediction networks, although some occasional timing errors at run time (Figure~\ref{fig:bad-example}) yet have to be fixed. A listening test on the NITech singing database showed that the proposed method scores similar to the state-of-art results by Ardaillon \etal\cite{ardaillon-2015} with the advantage of being data-driven.

Following this study we plan to verify the results by re-conducting the listening test on more singers. The next step is to incorporate other expression-related parameters such as loudness and vocal efforts in the transition-sustain framework. It is also interesting to apply the back-propagation based offset correction technique on the more general task of DNN-based text-to-speech synthesis.

\bibliographystyle{IEEEtran}

\bibliography{mybib}

\end{document}

%% file: boxplot.tikz
\begin{tikzpicture}

\begin{axis}[%
width=2.6in,
height=2in,
at={(0.4in,0.5in)},
scale only axis,
xmin=0.5,
xmax=3.5,
xtick={1,2,3},
xticklabels={{T-S Models},{HMM \cite{oura-2010}},{Multi-Layer \cite{ardaillon-2015}}},
ymin=-4,
ymax=2.5,
ytick={-4, -3, -2, -1, 0, 1, 2},
ylabel near ticks,
ylabel style={font=\color{white!15!black}},
ylabel={Normalized Difference Grades},
axis background/.style={fill=white},
ymajorgrids
]
\addplot [color=blue, forget plot]
  table[row sep=crcr]{%
0.8 -0.935271292455506\\
0.8 -0.807418639600256\\
0.8 -0.173512517280081\\
1.2 -0.173512517280081\\
1.2 -0.807418639600256\\
1.2 -0.935271292455506\\
1.2 -1.06312394531076\\
1.2 -1.58400481427217\\
0.8 -1.58400481427217\\
0.8 -1.06312394531076\\
0.8 -0.935271292455506\\
};
\addplot [color=blue, forget plot]
  table[row sep=crcr]{%
1.8 -1.72063201969703\\
1.8 -1.61923926090269\\
1.8 -1.15737855406235\\
2.2 -1.15737855406235\\
2.2 -1.61923926090269\\
2.2 -1.72063201969703\\
2.2 -1.82202477849136\\
2.2 -2.27596078177807\\
1.8 -2.27596078177807\\
1.8 -1.82202477849136\\
1.8 -1.72063201969703\\
};
\addplot [color=blue, forget plot]
  table[row sep=crcr]{%
2.8 -1.0041479321349\\
2.8 -0.879069337415218\\
2.8 -0.234111682151695\\
3.2 -0.234111682151695\\
3.2 -0.879069337415218\\
3.2 -1.0041479321349\\
3.2 -1.12922652685458\\
3.2 -1.61400009612496\\
2.8 -1.61400009612496\\
2.8 -1.12922652685458\\
2.8 -1.0041479321349\\
};
\addplot [color=black, forget plot]
  table[row sep=crcr]{%
1 -3.48038053388213\\
1 -1.58400481427217\\
};
\addplot [color=black, forget plot]
  table[row sep=crcr]{%
2 -3.8589170789706\\
2 -2.27596078177807\\
};
\addplot [color=black, forget plot]
  table[row sep=crcr]{%
3 -3.38603062127646\\
3 -1.61400009612496\\
};
\addplot [color=black, forget plot]
  table[row sep=crcr]{%
1 1.75051326587852\\
1 -0.173512517280081\\
};
\addplot [color=black, forget plot]
  table[row sep=crcr]{%
2 0.461749244828085\\
2 -1.15737855406235\\
};
\addplot [color=black, forget plot]
  table[row sep=crcr]{%
3 1.7101942123264\\
3 -0.234111682151695\\
};
\addplot [color=black, forget plot]
  table[row sep=crcr]{%
0.95  -3.48038053388213\\
1.05  -3.48038053388213\\
};
\addplot [color=black, forget plot]
  table[row sep=crcr]{%
1.95  -3.8589170789706\\
2.05  -3.8589170789706\\
};
\addplot [color=black, forget plot]
  table[row sep=crcr]{%
2.95  -3.38603062127646\\
3.05  -3.38603062127646\\
};
\addplot [color=black, forget plot]
  table[row sep=crcr]{%
0.95  1.75051326587852\\
1.05  1.75051326587852\\
};
\addplot [color=black, forget plot]
  table[row sep=crcr]{%
1.95  0.461749244828085\\
2.05  0.461749244828085\\
};
\addplot [color=black, forget plot]
  table[row sep=crcr]{%
2.95  1.7101942123264\\
3.05  1.7101942123264\\
};
\addplot [color=red, forget plot]
  table[row sep=crcr]{%
0.8 -0.935271292455506\\
1.2 -0.935271292455506\\
};
\addplot [color=red, forget plot]
  table[row sep=crcr]{%
1.8 -1.72063201969703\\
2.2 -1.72063201969703\\
};
\addplot [color=red, forget plot]
  table[row sep=crcr]{%
2.8 -1.0041479321349\\
3.2 -1.0041479321349\\
};
\addplot [color=red, draw=none, mark=+, mark options={solid, red}, forget plot]
  table[row sep=crcr]{%
1 2.05334957471471\\
1 2.90274520481948\\
2 1.34524744210202\\
2 1.24797893465185\\
2 1.03065523442495\\
2 1.08110146476311\\
2 1.08110146476311\\
2 1.08110146476311\\
2 0.919974116999838\\
2 1.99692284144571\\
2 0.815562844945955\\
2 0.815562844945955\\
2 1.00714124608352\\
2 -4.2125061080085\\
3 -3.82585069711619\\
3 1.98367358399366\\
3 1.90900889365266\\
3 1.90900889365266\\
3 -4.68315399783522\\
3 -4.21345255780028\\
};
\addplot [color=red, draw=none, mark=o, mark options={solid, red}, forget plot]
  table[row sep=crcr]{%
2 2.87627457102087\\
};
\end{axis}
\end{tikzpicture}%